\documentclass[11pt,preprint,aps,showpacs]{revtex4-1}
\usepackage{mathrsfs}
\usepackage{amsmath}
\usepackage{amssymb}
\usepackage{epsfig}
\usepackage{graphicx}
\usepackage{booktabs}
\usepackage{array}
\usepackage{multirow}
\usepackage{color}
\usepackage[colorlinks=true,linkcolor=red,citecolor=blue]{hyperref}
\begin{document}
\renewcommand{\arraystretch}{0.5}
\newcommand{\beq}{\begin{eqnarray}}
\newcommand{\eeq}{\end{eqnarray}}
\newcommand{\non}{\nonumber\\ }
\newcommand{\acp}{ {\cal A}_{CP} }
\newcommand{\psl}{ p \hspace{-1.8truemm}/ }
\newcommand{\nsl}{ n \hspace{-2.2truemm}/ }
\newcommand{\vsl}{ v \hspace{-2.2truemm}/ }
\newcommand{\epsl}{\epsilon \hspace{-1.8truemm}/\,  }
\newcommand{\tf}{\textbf}
\def \cpl{ Chin. Phys. Lett.  }
\def \ctp{ Commun. Theor. Phys.  }
\def \epjc{ Eur. Phys. J. C }
\def \jpg{  J. Phys. G }
\def \npb{  Nucl. Phys. B }
\def \plb{  Phys. Lett. B }
\def \prd{  Phys. Rev. D }
\def \prl{  Phys. Rev. Lett.  }
\def \zpc{  Z. Phys. C }
\def \jhep{ J. High Energy Phys.  }

%%%%%%%%%%%%%%%%%%%%%%%%%%%%%%%%%%%%%%%%%%%%%%%%%%%%
%%
\title{Cabibbo-Kobayashi-Maskawa-favored $B$ decays to a scalar meson and a $D$ meson}
\author{
Zhi-Tian Zou$^1$ \footnote{zouzt@ytu.edu.cn }, 
Ying Li$^1$ \footnote{liying@ytu.edu.cn}, 
Xin Liu$^2$ \footnote{liuxin@jsnu.edu.cn}
}
\affiliation
{\small
1.~Department of Physics, Yantai University, Yantai 264005,China\\
2.~School of Physics and Electronic Engineering, Jiangsu Normal University, Xuzhou 221116, China
}
\begin{abstract}
Within the perturbative QCD approach, we investigated the Cabibbo-Kobayashi-Maskawa-favored $B \to \overline{D} S$ (``$S$'' denoting the scalar meson) decays on the basis of the two-quark picture. Supposing the scalar mesons are the ground states or the first excited states, we calculated the branching ratios of 72 decay modes. Most of the branching ratios are in the range $10^{-4}$ to $10^{-7}$, which can be tested in the ongoing LHCb experiment and the forthcoming Belle-II experiment. Some decays, such as $B^+ \to \overline{D}^{(*)0} a_0^+(980/1450)$ and $B^+ \to D^{(*)-} a_0^+(980/1450)$, could be used to probe the inner structure and the nature of the scalar mesons, if the experiments are available. In addition, the ratios between the $Br(B^0\to \overline{D}^{(*)0}\sigma)$ and $Br(B^0\to \overline{D}^{(*)0}f_0(980))$ provide a potential way to determine the mixing angle between $\sigma$ and $f_0(980)$. Moreover, since in the standard model these decays occur only  through tree operators and  have no $CP$ asymmetries,  any deviation will be signal of the new physics beyond the standard model.
\end{abstract}
\maketitle
%=============================================================================
\section{Introduction}
%=============================================================================
Even though the quark-antiquark model works well for the pseudoscalar mesons and vector mesons, the study on the inner substructure of the scalar mesons is quite non-trivial, because the conventional quark-antiquark model cannot explain the properties of the scalar mesons below 1 GeV such as their light and inverted mass spectrum. Therefore, the understanding of the internal structure of the scalar mesons is one of the most interesting topics in hadron physics. Irrespective of the existence of $\sigma$ and $\kappa$ mesons, in the literatures, the scalar mesons have been identified as ordinary $\bar{q}q$ states, four-quark states or meson-meson bound states or even those supplemented with a scalar glueball. Unfortunately, we have not obtained a definite conclusion yet till now, due to the unknown nonperturbative properties of QCD. In hadron physics, most studies of the light scalar mesons are concentrated on the decay property of the scalar mesons and the production of the scalar mesons in $p\overline{p}$ (or $np$) collisions or the $\phi$ radiative decays \cite{prd016009}. After the first $B$ decay into a scalar meson, $B\to f_0(980)K$, was observed by Belle \cite{belle-f0980} and confirmed by BABAR \cite{babar-f0980}, the studies of the scalar mesons through hadronic $B$ decays have attracted more attentions because of the large phase space of $B$ decays.

The scalar mesons reported by experiments include the isosinglet $f_0(600)$($\sigma$), $f_0(980)$, $f_0(1370)$, $f_0(1500)$/$f_0(1710)$, the isodoublet $K_{0}^{*}(800)$($\kappa$) and $K_{0}^{*}(1430)$, and the isovector $a_0(980)$ and $a_0(1450)$. It is suggested that the scalar mesons with the mass below 1 GeV constitute one nonet, while those near 1.5 GeV form another one \cite{nonet,508,509,zheng1,zheng2}. As mentioned above, the inner structures of the scalar mesons are still unclear, though much effort has been devoted to interpreting the quark contents of the scalar mesons \cite{yangmz,try}. Now, it is accepted by most of us that the scalar meson above 1 GeV  can be identified as the traditional $q\overline{q}$ nonet with some possible glue content. However, the quark structure of the light scalar mesons below or near 1 GeV has been quite controversial. For example, $f_0(980)$  has been treated as a traditional $q\overline{q}$ state \cite{f0980-1}, as a four-quark $qq\overline{q}\overline{q}$ state \cite{f0980-2}, and even as a bound state of hadrons \cite{f0980-3}. The observation about the $D_s \to f_0(980) \pi^{+}$ decay introduces the probability of the $s\overline{s}$ component of $f_0(980)$, while $\Gamma(J/\psi \to f_0(980)\omega)\sim \Gamma(J/\psi \to f_0(980)\phi)$ indicates the existence of the non-strange components \cite{f0980-4-1,f0980-4-2}. Therefore, the isoscalars $f_0(980)$ and $f_0(600)$ perhaps have a mixing like the $\eta-\eta^{\prime}$ system. In the literatures, according to the category that the light mesons belong to, there are two typical scenarios for describing the scalar mesons. The scenario-1 (S1) is the naive 2-quark model: the nonet mesons below 1 GeV, such as $\kappa$, $a_0(980)$, $f_0(980)$, and $\sigma$, are treated as the lowest lying states, and these near 1.5 GeV, such as $a_0(1450)$, $K_0(1430)$, $f_0(1370/1500)$, are the first orbitally excited states. In scenario-2 (S2),  the nonet mesons near 1.5 GeV are viewed as the lowest lying states, while the mesons below 1 GeV may be the exotic states beyond the quark model such as four-quark bound states.

Recently, the LHCb collaboration have reported the measurements of the decays $B_{(s)} \to \overline{D} f_0(980)$ and $\overline{D}\sigma$ \cite{lhcb},
\begin{eqnarray}
&&Br(B^0 \to \overline{D}^0 \sigma)=(11.2\pm0.8\pm0.5\pm2.1\pm0.5)\times10^{-5},\nonumber\\
&&Br(B^0 \to \overline{D}^0 f_0(980))=(1.34\pm0.25\pm0.10\pm0.46\pm0.06)\times 10^{-5},\nonumber\\
&&Br(B^0_s \to \overline{D}^0 f_0(980))=(1.7\pm1.0\pm0.5\pm0.1)\times 10^{-6},
\label{valueexp}
\end{eqnarray}
which will shed light on the inner structure of the $\sigma$ and $f_0(980)$. On the theoretical side, these decays have been studied in ref.\cite{prd93} by using the perturbative QCD (PQCD) approach. To make clear the structure of the scalar mesons, more experiments and the systematical theoretical studies are both required. Experimentally, with more progressive accelerator and detector techniques, more and more two-body charmed $B$ decays with a light scalar meson in final states will be observed by LHCb experiment, as well as the forthcoming Belle-II experiments in future, which can provide more opportunities to explore the information about the scalar mesons. Theoretically, it is important  to study the charmed $B$ decays with scalars comprehensively. Motivated by this,  we shall study  the $B_{(s)} \to \overline{D}^{(*)}_{(s)} S$ decays in this work systematically for the first time. Together with these charmless $B$ decays with a light scalar meson \cite{Diehl:2001xe, qcdf1, qcdf2, qcdf3, qcdf4, pqcd1, pqcd2, pqcd3, pqcd4, pqcd5, pqcd6, pqcd7, pqcd8, pqcd9, pqcd10, pqcd11, pqcd12, pqcd13, pqcd14, pqcd15, Colangelo}, our studies will provide another insight into the flavor structures of the scalar mesons.

PQCD approach, based on the $k_T$ factorization, have been employed to study the two-body charmed $B$ decays, such as $B \to DP, DV, DA, DT$ decays \cite{dpv,da,dt}, where $P, V, A, T$ denote the pseudoscalar, vector, axial-vector, and tensor mesons, respectively. Most of the predictions are in good agreement with the present experimental data. Therefore, it is expected to be reliable to study the two-body charmed $B$ decays with a light scalar meson in final states. Very recently, in ref.\cite{ds}, we have studied the $B\to D^{(*)}S$ decays induced by $\overline{b} \to \overline{u}$ transition within PQCD approach, which are suppressed by the Cabibbo-Kobayashi-Maskawa (CKM) matrix element $|V_{ub}|$, but evade the suppression by the vector decay constants of the scalar mesons. In this work, we continue to study the CKM favored $B \to \overline{D}^{(*)} S$ decays with $\overline{b}\to \overline{c}$ transition, which are enhanced by the CKM matrix elements $|V_{cb}/V_{ub}|^2$, compared with the $B\to D^{(*)}S$ decays,  especially for these without strange quark in the four-quark operators. Compared with $B\to D^{(*)}S$ decays, for some $B \to \overline{D}^{(*)}S$ decays, the factorizable amplitude will vanish or be heavily suppressed due to the vanished and/or tiny vector decay constants of the scalar mesons, however, the hard-scattering emission diagrams and annihilation type diagrams perhaps provide sizable contributions, which is similar to the $B\to \overline{D}^{(*)}P, V, T$ decays. We thus expect that the branching ratios of some decays are large enough to be measured in the current LHC experiment and/or the forthcoming Belle-II in the future. It is worth pointing out that the annihilation type diagrams can be perturbatively calculated in the PQCD approach without endpoint singularity, and the PQCD approach has predicted the pure annihilation type decay modes successfully, such as the $B_s \to \pi^{+}\pi^{-}$ and $B^0\to D_s^- K^+$ \cite{bs,dpv,bdk}.

This paper is organized as follows: we will give a brief review of the formalism of the PQCD approach and specify the mesons' wave functions of the initial and final states in Sec.II. The perturbative calculations and the analytic formulas for the considered decays are given in Sec.III. The numerical results and phenomenological discussions will be presented in Sec.IV. The final section is reserved for summary.

%============================================================================
\section{FORMALISM AND WAVE FUNCTION}\label{sec:function}
As aforementioned, based on the $k_T$ factorization \cite{kt1,kt2,kt3},  PQCD approach  can effectively avoid the singularity by keeping the intrinsic transverse momenta of inner quarks. The kept transverse momenta will introduce the additional energy scale and lead to the double logarithms appearing the QCD radiative corrections, which can be resummed  to the Sudakov factor. As a result, the Sudakov factor will suppress the end-point region contribution and make  the calculation in the PQCD approach reliable and consistent.

The $B\to \overline{D}^{(*)} S$ decays only occur through tree operators and are governed by the effective Hamiltonian $H_{eff}$ \cite{heff}
\begin{eqnarray}
H_{eff}=\frac{G_F}{\sqrt{2}}V^*_{cb}V_{ud(s)}[C_1(\mu)O_1(\mu)+C_2(\mu)O_2(\mu)],
\end{eqnarray}
with the four-quark operators
\begin{eqnarray}
O_1=(\overline{b}_{\alpha} c_{\beta})_{V-A} (\overline{u}_{\beta} d(s)_{\alpha})_{V-A},\\
O_2=(\overline{b}_{\alpha}c_{\alpha})_{V-A} (\overline{u}_{\beta}d(s)_{\beta})_{V-A},
\end{eqnarray}
where  $\alpha$ and $\beta$ are the color indices, and $(\overline{b}_{\alpha}  c_{\beta})_{V-A}=\overline{b}_{\alpha}\gamma^{\mu}(1-\gamma^{5}) c_{\beta}$.
$V_{cb}$ and  $V_{ud(s)}$ are the CKM matrix elements. $C_{1,2}$ are the so-called Wilson coefficients at renormalization scale $\mu$.

To deal with the hadronic $B$ decays with multiple scales, the factorization hypothesis is usually adopted. The physics higher than the scale of the $W$ meson mass $(m_W)$ is electroweak and can be calculated perturbatively. Using the renormalization group techniques, we can evaluate the dynamical effects and get the Wilson coefficients from the $m_W$ scale to the $b$ quark mass ($m_b$) scale. The physics between $m_b$ scale and the factorization scale ($t$) can be calculated perturbatively, which is the so-called hard kernel in the PQCD approach. The dynamics below the factorizable scale is soft and nonperturbative but universal, which can be described by the hadronic wave functions of the mesons involving in the decays. According to facotrization above, in PQCD approach, the decay amplitudes can be written as the convolution of the Wilson coefficients $C(t)$, the hard kernel $H(x_i,b_i,t)$, and the hadronic wave functions $\Phi_{B,D,S}(x_i,b_i)$\cite{amp},
\begin{eqnarray}
\mathcal{A}\sim\int dx_1 dx_2 dx_3 b_1 db_1 b_2 db_2 b_3 db_3 \times \mathrm{Tr}[C(t)\Phi_B(x_1,b_1)\nonumber\\
\times\Phi_{D }(x_2,b_2)\Phi_{S }(x_3,b_3) H(x_i,,b_i,t)S_t(x_i)e^{-S(t)},
\label{eq:amplitude}
\end{eqnarray}
where $\mathrm{Tr}$ denotes the trace over Dirac and color indices, the $x_i(i=1,2,3)$ and $b_i$ are the longitudinal momentum fractions and conjugate variables of $k_{Ti}$ of the valence quarks in each meson, respectively. The threshold resummation of the double logarithms $\ln^2x_i$ lead to the jet function $S_t(x_i)$, which can smear the end-point singularity effectively \cite{jet}. The  aforementioned Sudakov factor $e^{-S(t)}$, coming from the resummation of the double logarithms $\ln^2(M_B/k_T)$, can suppress the soft dynamics effectively, i.e. the long distance contributions in the small $k_T$ region \cite{sudakov1,sudakov2}.

In order to provide reliable predictions in PQCD approach, the proper wave functions of initial and final states are essential. For the scalar mesons, the wave function can be defined as
\begin{eqnarray}
\Phi_S(x)=\frac{i}{2\sqrt{6}}[\makebox[-1.5pt][l]{/}p\phi_S(x)+m_S\phi_S^S(x)+m_S(\makebox[-1.5pt][l]{/}n\makebox[-1.5pt][l]{/}v-1)\phi_S^T(x)],
\end{eqnarray}
with the lightlike vectors $n=(1,0,\textbf{0}_T)$ and $v=(0,1,\textbf{0}_T)$. $\phi_S$ and $\phi_S^{S,T}$ are the leading-twist and twist-3 light-cone distribution amplitudes respectively, where $x$ is the momentum fraction of the ``quark". The leading twist light-cone distribution amplitude $\phi_S(x,\mu)$ of the scalar meson has the  general form \cite{zheng1,zheng2}
\begin{eqnarray}
\phi_S(x,\mu)=\frac{3}{2\sqrt{6}}x(1-x)[f_S(\mu)+\overline{f}_S\sum_{m=1}^{\infty}B_m(\mu)C_m^{3/2}(2x-1)],
\end{eqnarray}
with the Gegenbauer moments $B_m$ and the Gegenbauer polynomials $C_m^{3/2}$. For the twist-3 distribution amplitudes, we adopt the asymptotic forms for simplicity,
\begin{eqnarray}
\phi_S^{S}=\frac{\overline{f}_S}{2\sqrt{6}},\;\;\phi_S^T=\frac{\overline{f}_S}{2\sqrt{6}}(1-2x).
\end{eqnarray}
The $f_S$ and $\overline{f}_S$ are the vector decay constant and scalar decay constant of the scalar mesons, respectively. For the neutral scalar mesons, such as $\sigma$, $f_0$, and $a_0^0$, the vector decay constant vanishes required by the charge conjugation invariance or conservation of vector current. But the scalar decay constant $\overline{f}_S$, related by the equation
\begin{eqnarray}
\overline{f}_S=\mu f_S,\;\;\;
\mu=\frac{m_S}{m_2(\mu)-m_1(\mu)},
\end{eqnarray}
remains finite. Note that in different scenarios, the above parameters $B_m$, $f_S$, and $\overline{f}_S$ have different values,  which are referred to refs.\cite{zheng1,zheng2}.

For the $\sigma$ and $f_0(980)$, in the two-quark model, there exist so many experimental evidences to indicate the mixing between $\sigma$ and $f_0(980)$, which is like the mixing of the  $\eta-\eta^{\prime}$ system,
\begin{eqnarray}
\left(\begin{array}{c} \sigma\\
f_0 \end{array}\right)=\left(\begin{array}{cc}
\cos\theta &-\sin\theta\\
\sin\theta& \cos\theta\end{array}\right)\left(\begin{array}{c}f_n\\
f_s\end{array}\right),
\label{eq:f01}
\end{eqnarray}
with $f_n=(u\overline{u}+d\overline{d})/\sqrt{2}$ and $f_s=s\overline{s}$. For the mixing angle $\theta$, various experimental measurements have provided different values\cite{value}.
 Recently, the LHCb has proposed a upper limit $|\theta|<30^{\circ}$ by the process $\overline{B}^0\to J/\psi f_0(980)$ \cite{upper}. Analyzing the present experimental implications for the mixing angle, we prefer to adopt the two possible ranges of $25^{\circ}<\theta<40^{\circ}$ and $140^{\circ}<\theta<165^{\circ}$ \cite{angle}. It is noted that, for the $f_0(980)$ and $\sigma$ mesons, there are the other interpretations, for example, the $\pi \pi$ generalized distribution amplitudes \cite{pipi}.

For the $f_0(1370)-f_0(1500)$ system, according to ref.\cite{heavymix}, neglecting the tiny contributions from scalar glueball, the mixing form can be simplified as
\begin{eqnarray}
f_0(1370)&=&0.78f_n+0.51f_s,\nonumber\\
f_0(1500)&=&-0.54f_n+0.84f_s.
\label{eq:f02}
\end{eqnarray}

For the initial $B$ meson, neglecting the numerically suppressed Lorentz structure, the remained leading order wave function can be decomposed as \cite{Bwave}
\begin{eqnarray}
\Phi_B(x,b)=\frac{i}{\sqrt{6}}[(\makebox[-1.5pt][l]{/}P+m_B)\gamma_5\phi_B(x,b)].
\end{eqnarray}
The light-cone distribution amplitude $\phi_B(x,b)$ can be written as \cite{Bwave,Bamplitude}
\begin{eqnarray}
\phi_B(x,b)=N_Bx^2(1-x^2)\exp\left[-\frac{m_B^2x^2}{2\omega}-\frac{1}{2}\omega^2b^2\right],
\end{eqnarray}
with the normalization constant $N_B$, which can be determined through the following normalization condition
\begin{eqnarray}
\int _0^1 dx \phi_B(x,b=0)=\frac{f_B}{2\sqrt{6}}.
\end{eqnarray}
For the shape parameter $\omega$ and the decay constant $f_B$, we will take $(0.4\pm0.04)$GeV and $(0.19\pm0.02)$ GeV for the $B$ meson, respectively, and take $(0.5\pm0.05)$ GeV and $(0.23\pm0.03)$ GeV for the $B_s$ meson, due to the SU(3) breaking effects \cite{kt1,wfb1,wfb2}.

In terms of the heavy quark limit, the two-parton light cone distribution amplitudes of $D(D^*)$ meson will be taken as \cite{Dwave1,Dwave2,Dwave3,Dwave4}
\begin{eqnarray}
\langle D(p)|q_{\alpha}(z)\overline{c}_{\beta}(0)|0\rangle&=&\frac{i}{2\sqrt{6}}\int_0^1 dx e^{ixp\cdot z}[\gamma_5(\makebox[-1.5pt][l]{/}p +m_D)\phi_D(x,b)]_{\alpha,\beta},\\
\langle D^*(p)|q_{\alpha}(z)\overline{c}_{\beta}(0)|0\rangle&=&\frac{-1}{2\sqrt{6}}\int_0^1 dx e^{ixp\cdot z}[\makebox[-1.5pt][l]{/}\epsilon_L(\makebox[-1.5pt][l]{/}p + m_{D^*})\phi_{D^*}^L(x,b)\nonumber\\
&&+\makebox[-1.5pt][l]{/}\epsilon_T(\makebox[-1.5pt][l]{/}p + m_{D^*})\phi_{D^*}^T(x,b)]_{\alpha,\beta},
\end{eqnarray}
with the distribution amplitudes \cite{Dwave2,Dwave3,Dwave4}
\begin{eqnarray}
\phi_D(x,b)=\phi_{D^*}^{L,T}(x,b)=\frac{1}{2\sqrt{6}}f_{D^{(*)}}6x(1-x)[1+C_D(1-2x)]\exp[-\frac{1}{2}\omega^2b^2].
\end{eqnarray}
We choose $C_D=0.5\pm0.1$, $\omega=0.1$ GeV and $f_D=207$ MeV for the $D$ meson,  and $C_D=0.4\pm0.1$, $\omega=0.2$ GeV and $f_{D_s}=241$ MeV for the $D_s$ meson \cite{Dparameter}. For $D_{(s)}^*$, the decay constants can be obtained through the relation based on the heavy qurak effective theory, which can be found in refs.\cite{dpv,dt}.

%============================================================================
\section{PERTURBATIVE CALCULATION}\label{sec:amplitude}

\begin{figure}[!tbp]
\begin{center}
\vspace{-6cm}
\centerline{\epsfxsize=10 cm \epsffile{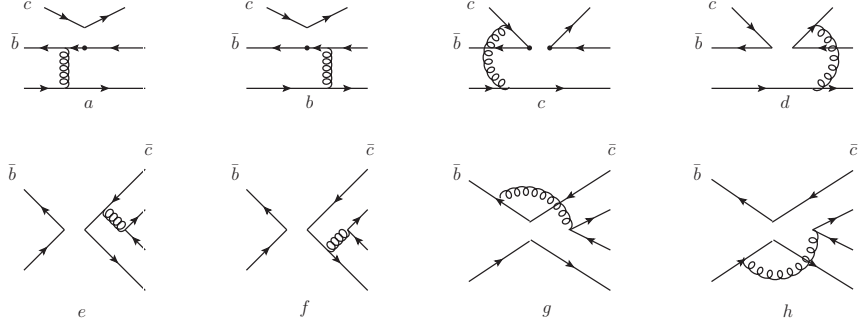}}
\vspace{-4.5cm}
\caption{Leading order Feynman diagrams contributing to the
$B\,\rightarrow\,\overline{D}^{(*)}S$ decays in PQCD appraoch}
\label{fig:diagram1}
 \end{center}
\end{figure}

\begin{figure}[!tbp]
\begin{center}

\vspace{-3cm}
\centerline{\epsfxsize=11cm \epsffile{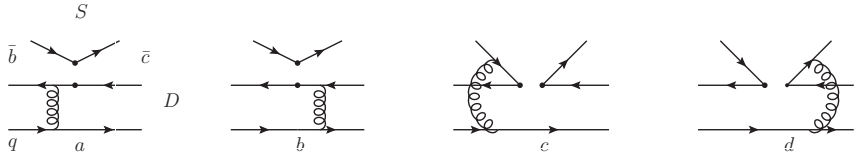}}
\vspace{-11.5cm}
\caption{Leading order Feynman diagrams contributing to the
$B\,\rightarrow\,\overline{D}^{(*)}S$ decays with a scalar meson emitted in PQCD}
\label{fig:diagram2}

\end{center}
\end{figure}

In this section, within the PQCD approach, we specifically calculate the decay amplitudes without the Wilson coefficients in eq.(\ref{eq:amplitude}) for each Feynman diagram, and express the calculated amplitudes as the convolution of the hard kernel and the mesons' wave functions. It is noted that, there are two kinds of diagrams contributing to the considered decays at the leading order.  The diagrams with  a $\overline{D}$ meson emitted are presented in Fig.\ref{fig:diagram1}, and those with a scalar meson emitted are listed in Fig.\ref{fig:diagram2}.

For the two factorizable emission diagrams (a) and (b) in Fig.\ref{fig:diagram1}, the amplitudes can be written as
\begin{eqnarray}
\mathcal{A}_{ef}&=&8\pi C_f f_D m_B^4\int_0^1dx_1dx_2\int_0^{1/\Lambda}b_1db_1b_3db_3\phi_B(x_1,b_1)\nonumber\\
&\times &\{[\phi_S(x_3)(r_D^2(2x_3+1)-(x_3+1))+r_S(2x_3-1)(\phi_S^S(x_3)+\phi_S^T(x_3))]\nonumber\\
&& \cdot E_{ef}(t_a)h_{ef}(x_1,x_3(1-r_D^2),b_1,b_3)\nonumber\\
& &-2r_S\phi_S^S(x_3)E_{ef}(t_b)h_{ef}(x_3,x_1(1-r_D^2),b_3,b_1)\},
\label{eq.ef}
\end{eqnarray}
where $r_S=m_S/m_B$, $r_D=m_D/m_B$, and the color factor $C_f=4/3$ for $B$ decays. The expressions of the scale $t$, Sudakov factor $E$, and the hard functions $h$ from the denominator of the propagator can be found in Appendix.A of ref.\cite{dt}

The two diagrams (c) and (d) in Fig.\ref{fig:diagram1} are the so-called hard-scattering emission diagrams. Compared to the previous two, each decay amplitude involves three meson wave functions. After integrating out  $b_3$ with $\delta$ function $\delta(b_1-b_3)$, the amplitudes for these two diagrams can be expressed by
 \begin{eqnarray}
 \mathcal{M}_{enf}&=&16\sqrt{\frac{2}{3}}\pi C_f m_B^4\int_0^1dx_1dx_2dx_3\int_0^{1/\Lambda}b_1db_1b_2db_2\phi_B(x_1,b_1)\phi_D(x_2,b_2)\nonumber\\
 &\times&\{[\phi_S(x_3)(r_D^2-r_S^2(2x_2+x_3-2)+x_2-1)\nonumber\\
 &&+r_Sx_3(\phi_S^S(x_3)-\phi_S^T(x_3))]E_{enf}(t_c)h_{enf1}(x_i,b_i)\nonumber\\
 & &-[\phi_S(x_3)(x_2(r_D^2+2r_S^2-1)+x_3(2r_D^2+r_S^2-1))\nonumber\\
 & & +r_Sx_3(\phi_S^S(x_3)+\phi_S^T(x_3))]E_{enf}(t_d)h_{enf2}(x_i,b_i)\}.
 \label{eq.enf}
 \end{eqnarray}
The four diagrams in the second row are the annihilation type diagrams, which can be perturbatively calculated in the PQCD approach.  (e) and (f) are the factorizable annihilation diagrams with the $B$ meson factorized out. The amplitudes can be written as:
\begin{eqnarray}
\mathcal{A}_{af}&=&8\pi C_f f_B m_B^4\int_0^1dx_2dx_3\int_0^{1/\Lambda}b_2db_2b_2db_3\phi_D(x_2,b_2)\nonumber\\
&\times&\{[\phi_S(x_3)(r_D^2(2x_3-3)+(r_S^2-1)(x_3-1))+r_Dr_S(\phi_S^S(x_3)(2x_3-3)\nonumber\\
&&-\phi_S^T(x_3)(2x_3-1)]E_{af}(t_e)h_{af}((1-x_3),x_2(1-r_D^2),b_2,b_3)\nonumber\\
&&+[\phi_S(x_3)((r_D^2-1)x_2+r_S^2(2x_2-1))+2r_Dr_S(x_2+1)\phi_S^S(x_3)]\nonumber\\
&&\cdot E_{af}(t_f)h_{af}(x_2,(1-x_3)(1-r_D^2),b_3,b_2)\}.
\label{af}
\end{eqnarray}
For the nonfactorizable annihilation diagrams (g) and (h), the corresponding amplitudes are as following:
\begin{eqnarray}
\mathcal{M}_{anf}&=&-16\sqrt{\frac{2}{3}}\pi C_f m_B^4\int_0^1dx_1dx_2dx_3\int_0^{1/\Lambda}b_1db_1b_2db_2 \phi_B(x_1,b_1)\phi_D(x_3,b_2)\nonumber\\
&&\times\{[\phi_S(x_3)(r_D^2+r_S^2(2x_2+x_3-1)-x_2)+r_Dr_S(\phi_S^S(x_3)(x_2-x_3+3)\nonumber\\
&&+\phi_S^T(x_3)(1-x_2-x_3))]E_{anf}(t_g)h_g(x_i,b_1,b_2)\nonumber\\
&&+[\phi_S(x_3)(r_D^2(x_2+2x_3-2)-x_3+1)-r_Dr_S(\phi_S^S(x_3)(x_2-x_3+1)\nonumber\\
&&+\phi_S^T(x_3)(x_2+x_3-1))]E_{anf}(t_h)h_h(x_i,b_1,b_2).\label{anf}
\end{eqnarray}

For these diagrams with a scalar meson emitted in Fig.\ref{fig:diagram2}, the decay amplitudes are expressed as:
\begin{eqnarray}
\mathcal{A}_{ef}^{\prime}&=&8\pi C_f f_S m_B^4 \int_0^1 dx_1dx_3\int_0^{1/\Lambda}b_1db_1b_3db_3\phi_B(x_1,b_1)\phi_D(x_3,b_3) \nonumber\\
&\times&\{[r_D(2x_3-1)-(1+x_3)]E_{ef}(t_a)h_{ef}(x_1,x_3,b_1,b_3)\nonumber\\
&&-rdE_{ef}(t_b)h_ef(x_3,x_1,b_3,b_1)\},
\label{eq:aefs}
\end{eqnarray}
\begin{eqnarray}
\mathcal{M}_{enf}^{\prime}&=&-16\sqrt{\frac{2}{3}}\pi C_f m_B^4\int_0^1dx_1dx_2dx_3\int_0^{1/\Lambda}b_1db_1b_2db_2\phi_B(x_1,b_1)\phi_D(x_3,b_1)\phi_S(x_2)\nonumber\\
&\times&\{[r_D^2(2x_2+x_3-2)-r_Dx_3-x_2+1]E_{enf}(t_c^{\prime})h_{enf1}^{\prime}(x_i,b_i)\nonumber\\
&&+[x_2(2r_D^2+r_S^2-1)+x_3(r_D^2+r_D+2r_S^2-1)]E_{enf}(t_d^{\prime})h_{enf2}^{\prime}(x_i,b_i)\}.
\label{eq:enfs}
\end{eqnarray}
From the eq.(\ref{eq:aefs}), one can find that, the factorizable emission diagrams with a scalar meson emitted are highly suppressed by the vector decay constant of the scalar meson, and even vanish for the decays emitting a neutral scalar meson because the neutral scalar meson cannot be produced through  $(V-A)$ current.

For the $B\to \overline{D}^*S$ decays, there are only the longitudinal polarization contributions required by the conservation of angular momentum.  The expressions of the factorizable emission  contributions can be obtained by the following substitutions in eq.(\ref{eq.ef}):
\begin{eqnarray}
\mathcal{A}\to -\mathcal{A}^L\;\;,\phi_D\,\to\,\phi^{L} _{D^*},\;\;\; m_D\,\to\,m_{D^*},\;\;\;f_D\to f_{D^*}^L.
\label{eq:substitution}
\end{eqnarray}
For the hard-scattering emission diagrams, the decay amplitudes can be expressed as:
\begin{eqnarray}
\mathcal{M}_{enf}^L&=&16\sqrt{\frac{2}{3}}\pi C_f m_B^4\int_0^1dx_1dx_2dx_3\int_0^{1/\Lambda }b_1db_1b_2db_2\phi_B(x_1,b_1)\phi_{D^*}^L(x_2,b_2)\nonumber\\
&\times& \Big\{\Big[\phi_S(x_3)(r_D^2(1-2x_2)+r_S^2(x_2+x_3-1)+(\frac{1}{2}r_S^2-1)(x_2-1))\nonumber\\
&&-r_Sx_3(\phi_S^S(x_3)-\phi_S^T(x_3))\Big]E_{enf}(t_c)h_{enf1}(x_i,b_i)\nonumber\\
&&+\Big[\phi_S(x_3)(x_2(r_D^2+\frac{3}{2}r_S^2-1)+x_3(2r_D^2+\frac{1}{2}r_S^2-1))\nonumber\\&&+r_Sx_3(\phi_S^S(x_3)+\phi_S^T(x_3))\Big]E_{enf}(t_d)h_{enf2}(x_i,b_i)\Big\}.
\end{eqnarray}
The annihilation type contributions can be written as:
\begin{eqnarray}
\mathcal{A}_{af}^L&=&8\pi C_f f_B m_B^4\int_0^1dx_2dx_3\int_0^{1/\Lambda}b_2db_2b_3db_3\phi^L_{D^*}(x_2,b_2)\nonumber\\
&\times&\Big\{\Big[\phi_S(x_3)(r_D^2(1-2x_3)+(1-\frac{1}{2}r_S^2)(x_3-1))\nonumber\\
&&+r_Dr_S(\phi_S^S(x_3)+\phi_S^T(x_3))\Big] h_{af}((1-x_3),x_2(1-r_D^2),b_2,b_3)E_{af}(t_e)\nonumber\\
&&+\Big[\phi_S(x_3)((1-rd_D^2)x_2+r_S^2(1-\frac{3}{2}x_2))\nonumber\\
&&+2r_Dr_S(1-x_2)\phi_S^S(x_3)\Big]h_{af}(x_2,(1-x_3)(1-r_D^2),b_3,b_2)E_{af}(t_f)\Big\},
\label{afl}
\end{eqnarray}
\begin{eqnarray}
\mathcal{M}_{anf}^L&=&-16\sqrt{\frac{2}{3}}\pi C_f m_B^4\int_0^1dx_1dx_2dx_3\int_0^{1/\Lambda}b_1db_1b_2db_2\phi_B(x_1,b_1)\phi_{D^*}^L(x_2,b_2)\nonumber\\
&\times&\Big\{\Big[\phi_S(x_3)(r_D^2(1-2x_2)-r_S^2(x_2+x_3-1)+x_2(1-\frac{1}{2}r_S^2))\nonumber\\
&&-r_Dr_S((x_2+x_3-1)\phi_S^S(x_3)+(1-x_2+x_3)\phi_S^T(x_3))\Big]h_g(x_{i},b_1,b_2)E_{anf}(t_g)\nonumber\\
&&+\Big[\phi_S(x_3)(r_D^2(x_2-2x_3+2)+(1+\frac{1}{2}r_S^2)(x_3-1))\nonumber\\
&&-r_Dr_S((x_2+x_3-1)\phi_S^S(x_3)+(x_2-x_3+1)\phi_S^T(x_3))\Big]h_h(x_{i},b_1,b_2)E_h(t_h)\Big\}.
\end{eqnarray}

For these diagrams with a scalar meson emitted in  $B\to \overline{D}^* S$ decays, the factorizable emission contributions can be obtained by adopting the same substitutions as the eq.(\ref{eq:substitution}) in eq.(\ref{eq:aefs}) , and the hard-scattering emission contributions can be expressed as:
\begin{eqnarray}
\mathcal{M}_{enf}^{\prime L}&=&16\sqrt{\frac{2}{3}}\pi C_f m_B^4\int_0^1dx_1dx_2dx_3\int_0^{1\Lambda}\phi_{B}(x_1,b_1)\phi_{D^*}^L(x_3,b_1)\phi_S(x_2)\nonumber\\
&\times&\Big\{\Big[r_D^2(x_2-x_3-2)+r_Dx_3+r_S^2x_2+(1-x_2)(1+\frac{1}{2}r_S^2)\Big]E_{enf}(t_c^{\prime})h_{enf1}^{\prime}(x_i,b_i)\nonumber\\
&&+\Big[x_2(r_D^2-1-\frac{1}{2}r_S^2)+x_3(r_D^2+r_D+\frac{3}{2}r_S^2-1)\Big]E_{enf}(t_d^{\prime})h_{enf2}^{\prime}(x_i,b_i)\Big\}.
\label{eq:enfsl}
\end{eqnarray}

The expressions of complete decay amplitudes with the Wilson coefficients are the same as those of $B_q \to \overline{D}^{(*)}_{(s)} T$ decays, which can be found in the Appendix B of ref.\cite{dt}, because the topologies of these two type decays are identical.
%============================================================================
\section{NUMERICAL RESULTS AND DISCUSSIONS}\label{sec:result}
%============================================================================
In this section, we will present the numerical results and give some phenomenological analyses on those considered $B_q\to \overline{D}_{(s)}^{(*)} S$ decays. At the beginning, we should list the input parameters in numerical calculations:
\begin{eqnarray}
 &&\Lambda_{\overline{MS}}^{f=4}=0.25\pm0.05 \mathrm{GeV},\;\;m_{B_{(s)}}=5.28(5.37)\mathrm{GeV},\;\;m_b=4.8\mathrm{GeV},\nonumber\\
 &&m_{D_{(s)}}=1.869/1.968\mathrm{GeV},\;\;m_{D_{(s)}^*}=2.010/2.112\mathrm{GeV},\nonumber\\
  &&\tau_{B^{\pm/0}}=1.641/1.519 ps,\;\;\;\tau_{B_{s}}=1.479 ps,\nonumber\\
  &&V_{cb}=0.0412_{-0.0005}^{+0.0011},\;\;V_{us}=0.22534\pm0.00065 ,\;\;V_{ud}=0.97427\pm0.00015 .
  \label{eq:parameter}
 \end{eqnarray}
For the decay constants of the scalar mesons, we adopt the same values as the ref. \cite{zheng2}.

\begin{table}[!ht]
\centering
\caption{Branching ratios of $B_{q} \to \overline{D}S(a_0(980),\kappa, \sigma, f_0(980))$ decays calculated in the PQCD approach in S1.}
\begin{tabular}[t]{l!{\;\;\;\;}c!{\;\;}c}
\hline\hline
\multirow{2}{*}{Decay Modes} & \multirow{2}{*}{Class} & \multirow{2}{*}{BRs($10^{-6}$)} \\
  &&\\
 \hline
 \vspace{0.05cm}
$B^{+}\to \overline{D}^{0}a_{0}^+$&C&$483_{-215-59-12}^{+244+52+26}$\\
\vspace{0.05cm}
$B^{0}\to D^{-}a_0^+$ &T&$17.6_{-7.6-7.5-0.4}^{+9.8+10.8+0.9}$\\
\vspace{0.05cm}
$B^{0}\to \overline{D}^{0}a_{0}^0$& C & $160_{-75-26-4}^{+88+28+9}$\\
\vspace{0.05cm}
$B^{0}\to \overline{D}^{0}\sigma (f_n)$ & C & $134_{-55-33-3}^{+65+28+7}$\\
\vspace{0.05cm}
$B^{0}\to \overline{D}^{0}f_0 (f_n)$ &C&$78.4_{-36.2-31.0-1.9}^{+42.8+28.2+4.3}$\\
\vspace{0.05cm}
$B^{0}\to D_s^{-}\kappa^{+}$&E&$69.2_{-21.2-9.2-1.7}^{+23.1+7.7+3.8}$ \\
\vspace{0.05cm}
$B_{s}\to \overline{D}^{0}\overline{\kappa}$&C&$262_{-131-53-7}^{+154+46+14}$\\
\vspace{0.05cm}
$B_s\to D_s^- a_0^+$&T&$61.2_{-27.2-9.2-1.5}^{+33.6+8.7+3.3}$\\

\hline
\vspace{0.05cm}
$B^{+}\to \overline{D}^{0} \kappa^{+}$&C&$10.8_{-6.4-2.0-0.3}^{+7.9+1.8+0.7}$\\
\vspace{0.05cm}
$B^{0}\to D^{-}\kappa^+$& T &$4.83_{-1.64-0.57-0.14}^{+1.92+0.57+0.29}$\\
\vspace{0.05cm}
$B^{0}\to \overline{D}^{0}\kappa$ &C&$6.89_{-4.28-2.09-0.21}^{+5.26+1.92+0.41}$\\
\vspace{0.05cm}
$B_{s}\to D^{-}a_{0}^{+}$&E&$3.42_{-1.17-0.44-0.10}^{+1.31+0.40+0.21}$\\
\vspace{0.05cm}
$B_{s}\to \overline{D}^{0}a_{0}$&E&$1.70_{-0.58-0.21-0.05}^{+0.65+0.21+0.10}$\\
\vspace{0.05cm}
$B_{s}\to \overline{D}^{0}\sigma (f_n)$&E&$1.13_{-0.40-0.14-0.04}^{+0.42+0.12+0.07}$\\
\vspace{0.05cm}
$B_{s}\to \overline{D}^{0}\sigma (f_s)$ &C&$14.2_{-6.7-2.2-0.4}^{+7.9+2.0+0.9}$\\
\vspace{0.05cm}
$B_{s}\to \overline{D}^{0}f_0 (f_n)$&E&$1.36_{-0.45-0.15-0.04}^{+0.51+0.15+0.08}$\\
\vspace{0.05cm}
$B_s \to \overline{D}^0 f_0 (f_s)$&C&$10.6_{-5.3-2.1-0.3}^{+6.1-2.0-0.6}$\\
\vspace{0.05cm}
$B_{s}\to D_{s}^{-}\kappa^{+}$&T&$0.94_{-0.43-0.24-0.03}^{+0.56+0.24+0.06}$\\
 \hline\hline
\end{tabular}\label{T1}
\end{table}

 \begin{table}[!ht]
\centering
 \caption{Branching ratios of $B_{(s)}\to \overline{D} S (a_0(1450), K_0^*(1430), f_0(1370)$, and $f_0(1500))$  calculated in the PQCD approach in S1 and S2, respectively.}
\begin{tabular}[t]{l!{\;\;\;\;}c!{\;\;}c!{\;\;}c}
\hline\hline
  \multirow{2}{*}{Decay Modes} & \multirow{2}{*}{Class} & \multirow{2}{*}{BRs($10^{-5}$)$S_1$}&\multirow{2}{*}{BRs($10^{-5}$)$S_2$} \\
  &&\\
 \hline
 \vspace{0.05cm}
$B^{+}\to \overline{D}^{0}a_{0}^+(1450)$&C&$72.1_{-31.6-11.5-1.8}^{+39.2+10.8+3.9}$&$123_{-64-14-4}^{+73+10+5}$\\
\vspace{0.05cm}
$B^{0}\to D^{-}a_0^+(1450)$ &T,E&$4.09_{-2.09-0.73-0.10}^{+2.63+1.43+0.23}$&$0.92_{-0.43-0.22-0.02}^{+0.57+0.38+0.05}$\\
\vspace{0.05cm}
$B^{0}\to \overline{D}^{0}a_0(1450)$&C&$31.3_{-14.3-5.5-0.8}^{+17.4+4.2+1.7}$&$66.2_{-32.8-7.1-1.7}^{+37.6+6.0+3.6}$\\
\vspace{0.05cm}
$B^{0}\to \overline{D}^{0}f_0(1370) (f_n)$ &C&$28.6_{-12.0-3.3-0.8}^{+13.7+2.1+1.5}$&$16.3_{-7.6-4.0-0.4}^{+9.1+4.4+0.8}$\\
\vspace{0.05cm}
$B^{0}\to \overline{D}^{0}f_{0}(1500) (f_n)$ &C&$27.2_{-111.1-2.6-0.7}^{+13.2+2.1+1.5}$&$13.2_{-5.9-2.6-0.3}^{+7.4+3.5+0.7}$\\
\vspace{0.05cm}
$B^{0}\to D_s^{-}K_{0}^{*+}(1430)$&E&$1.29_{-0.43-0.42-0.03}^{+0.47+0.38+0.07}$ &$8.60_{-3.56-0.49-0.21}^{+3.97+0.48+0.47}$\\
\vspace{0.05cm}
$B_{s}\to \overline{D}^{0}\overline{K}_0^{*}(1430)$ &C&$53.9_{-23.0-6.1-1.3}^{+26.8+3.1+2.9}$&$68.8_{-35.4-8.0-1.7}^{+38.3+6.3+3.8}$\\
\vspace{0.05cm}
$B_{s}\to D_s^{-}a_0^{+}$(1450)&T&$9.80_{-4.66-1.65-0.24}^{+5.57+1.66+0.54}$&$4.11_{-2.29-0.80-0.11}^{+3.09+0.86+0.22}$\\
\hline
\vspace{0.05cm}
$B^{+}\to \overline{D}^0 K_0^{*+}(1430)$&C&$4.72_{-1.83-0.87-0.15}^{+2.22+0.81+0.28}$&$4.96_{-3.15-0.34-0.15}^{+4.02+0.71+0.29}$\\
\vspace{0.05cm}
$B^{0}\to {D}^- K_0^{*+}(1430)$&T&$0.97_{-0.36-0.11-0.03}^{+0.45+0.12+0.06}$&$0.79_{-0.38-0.09-0.03}^{+0.46+0.11+0.05}$\\
\vspace{0.05cm}
$B^{0}\to \overline{D}^0 K_0^{*0}(1430)$&C&$3.39_{-1.27-0.45-0.10}^{+1.47+0.36+0.21}$&$3.19_{-2.34-0.40-0.09}^{+3.21+0.40+0.20}$\\
\vspace{0.05cm}
$B_{s}\to {D}^- a_0^{+} (1450)$&E&$0.14_{-0.07-0.01-0.01}^{+0.06+0.01+0.01}$&$0.43_{-0.18-0.03-0.01}^{+0.21+0.02+0.02}$\\
\vspace{0.05cm}
$B_{s}\to \overline{D}^0 a_0^{0}(1450) $&E&$0.07_{-0.03-0.01-0.01}^{+0.03+0.01+0.01}$&$0.21_{-0.09-0.01-0.01}^{+0.11+0.01+0.02}$\\
\vspace{0.05cm}
$B_{s}\to \overline{D}^0 f_0(1370)(f_n) $&E&$0.05_{-0.02-0.01-0.01}^{+0.04+0.01+0.01}$&$0.17_{-0.07-0.01-0.01}^{+0.09+0.01+0.01}$\\
\vspace{0.05cm}
$B_{s}\to \overline{D}^0 f_0(1370)(f_s) $&C&$2.30_{-1.12-0.25-0.07}^{+1.46+0.13+0.14}$&$2.97_{-1.53-0.31-0.09}^{+2.99+0.24+0.18}$\\
\vspace{0.05cm}
$B_{s}\to \overline{D}^0 f_0(1500)(f_n) $&E&$0.05_{-0.02-0.01-0.01}^{+0.04+0.01+0.01}$&$0.17_{-0.07-0.01-0.01}^{+0.10+0.01+0.01}$\\
\vspace{0.05cm}
$B_{s}\to \overline{D}^0 f_0(1500)(f_s) $&C&$2.24_{-1.08-0.22-0.07}^{+1.39+0.14+0.13}$&$2.71_{-2.04-0.24-0.08}^{+2.81+0.21+0.17}$\\
\vspace{0.05cm}
$B_{s}\to D_s^- K_0^{*+}(1430) $&T,E&$0.35_{-0.18-0.08-0.01}^{+0.24+0.12+0.02}$&$0.37_{-0.16-0.03-0.01}^{+0.19+0.02+0.02}$\\
\hline\hline
\end{tabular}\label{T2}
\end{table}

  \begin{table}[!ht]
\centering
 \caption{Branching ratios of $B_{q} \to \overline{D}^* S(a_0,\kappa, \sigma, f_0)$ decays calculated in the PQCD approach in S1.}
\begin{tabular}[t]{l!{\;\;\;\;}c!{\;\;}c}
\hline\hline
  \multirow{2}{*}{Decay Modes} & \multirow{2}{*}{Class} & \multirow{2}{*}{BRs($10^{-6}$)} \\
  &&\\
 \hline
 \vspace{0.05cm}
$B^{+}\to \overline{D}^{*0}a_{0}^+$&C&$520_{-188-126-13}^{+215+104+29}$\\
\vspace{0.05cm}
$B^{0}\to D^{*-}a_0^+$ &T&$250_{-84-64-7}^{+91+57+13}$\\
\vspace{0.05cm}
$B^{0}\to \overline{D}^{*0}a_{0}$& C & $128_{-67-34-3}^{+76+28+7}$\\
\vspace{0.05cm}
$B^{0}\to \overline{D}^{*0}\sigma (f_n)$ & C & $171_{-70-45-4}^{+78+36+9}$\\
\vspace{0.05cm}
$B^{0}\to \overline{D}^{*0}f_0 (f_n)$ &C&$119_{-51-43-4}^{+57+38+6}$\\
\vspace{0.05cm}
$B^{0}\to D_s^{*-}\kappa^{+}$&E&$12.3_{-4.1-2.4-0.3}^{+4.3+1.9+0.7}$ \\
\vspace{0.05cm}
$B_{s}\to \overline{D}^{*0}\overline{\kappa}$&C&$320_{-153-65-8}^{+178+56+17}$\\
\vspace{0.05cm}
$B_s\to D_s^{*-} a_0^+$&T&$162_{-62-49-4}^{+68+49+9}$\\
\hline
\vspace{0.05cm}
$B^{+}\to \overline{D}^{*0} \kappa^{+}$&C&$8.80_{-4.41-3.00-0.27}^{+5.78+2.51+0.53}$\\
\vspace{0.05cm}
$B^{0}\to D^{*-}\kappa^+$& T &$3.42_{-1.32-0.85-0.10}^{+1.58+1.10+0.21}$\\
\vspace{0.05cm}
$B^{0}\to \overline{D}^{*0}\kappa$ &C&$9.25_{-5.03-3.06-0.27}^{+6.54+2.52+0.56}$\\
\vspace{0.05cm}
$B_{s}\to D^{*-}a_{0}^{+}$&E&$0.76_{-0.28-0.14-0.03}^{+0.30+0.14+0.04}$\\
\vspace{0.05cm}
$B_{s}\to \overline{D}^{*0}a_{0}$&E&$0.38_{-0.15-0.07-0.02}^{+0.14+0.06+0.02}$\\
\vspace{0.05cm}
$B_{s}\to \overline{D}^{*0}\sigma (f_n)$&E&$0.19_{-0.08-0.04-0.01}^{+0.07+0.03+0.01}$\\
\vspace{0.05cm}
$B_{s}\to \overline{D}^{*0}\sigma (f_s)$ &C&$16.4_{-7.6-2.8-0.5}^{+8.8+2.4+1.0}$\\
\vspace{0.05cm}
$B_{s}\to \overline{D}^{*0}f_0 (f_n)$&E&$0.27_{-0.10-0.06-0.01}^{+0.10+0.04+0.01}$\\
\vspace{0.05cm}
$B_s \to \overline{D}^{*0} f_0 (f_s)$&C&$12.7_{-6.0-2.7-0.3}^{+7.1+2.6+0.8}$\\
\vspace{0.05cm}
$B_{s}\to D_{s}^{*-}\kappa^{+}$&T&$6.47_{-2.77-1.61-0.20}^{+3.23+1.71+0.39}$\\
 \hline\hline
\end{tabular}\label{T3}
\end{table}

 \begin{table}[!ht]
\centering
 \caption{Branching ratios of $B_{(s)}\to \overline{D}^* S (a_0(1450), K_0^*(1430), f_0(1370)$, and $f_0(1500))$  calculated in the PQCD approach in S1 and S2, respectively.}
\begin{tabular}[t]{l!{\;\;\;\;}c!{\;\;}c!{\;\;}c}
\hline\hline
  \multirow{2}{*}{Decay Modes} & \multirow{2}{*}{Class} & \multirow{2}{*}{BRs($10^{-5}$)($S_1$)} &\multirow{2}{*}{BRs($10^{-5}$)($S_2$)}\\
  &&\\
 \hline
 \vspace{0.05cm}
$B^{+}\to \overline{D}^{*0}a_{0}^+(1450)$&C&$207_{-81-20-5}^{+90+19+11}$&$98.2_{-56.1-26.8-2.5}^{+70.1+25.8+5.3}$\\
\vspace{0.05cm}
$B^{0}\to D^{*-}a_0^+(1450)$ &T &$26.8_{-11.1-7.1-0.6}^{+12.7+6.6+1.5}$&$11.3_{-5.9-2.6-0.2}^{+7.4+2.1+0.7}$\\
\vspace{0.05cm}
$B^{0}\to \overline{D}^{*0}a_0(1450)$&C &$40.1_{-17.3-5.0-1.0}^{+21.2+4.1+2.1}$&$58.9_{-31.2-8.4-1.4}^{+36.7+7.3+3.3}$\\
\vspace{0.05cm}
$B^{0}\to \overline{D}^{*0}f_0(1370) (f_n)$ &C&$44.8_{-18.4-4.0-1.1}^{+21.3+1.5+2.4}$&$27.8_{-21.4-5.6-0.7}^{+32.0+6.1+1.6}$\\
\vspace{0.05cm}
$B^{0}\to \overline{D}^{*0}f_{0}(1500) (f_n)$ &C&$44.5_{-18.1-3.3-1.0}^{+21.2+1.9+2.5}$&$25.1_{-20.1-4.5-0.6}^{+30.0+5.0+1.4}$\\
\vspace{0.05cm}
$B^{0}\to D_s^{*-}K_{0}^{*+}(1430)$&E&$0.50_{-0.16-0.11-0.01}^{+0.18+0.14+0.03}$ &$0.93_{-0.46-0.18-0.02}^{+0.54+0.23+0.06}$\\
\vspace{0.05cm}
$B_{s}\to \overline{D}^{*0}\overline{K}_0^{*}(1430)$ &C&$73.0_{-30.7-5.8-1.8}^{+36.5+3.0+4.0}$&$79.0_{-55.6-10.8-2.0}^{+72+8.0+4.3}$\\
\vspace{0.05cm}
$B_{s}\to D_s^{*-}a_0^{+}$(1450)&T&$21.9_{-9.6-7.1-0.5}^{+10.6+7.7+1.2}$&$5.91_{-3.43-1.55-0.15}^{+4.36+1.40+0.32}$\\
\hline
\vspace{0.05cm}
$B^{+}\to \overline{D}^{*0} K_0^{*+}(1430)$&C&$10.1_{-3.3-1.0-0.3}^{+3.9+1.0+0.6}$&$3.08_{-2.25-0.67-0.08}^{+3.62+0.73+0.18}$\\
\vspace{0.05cm}
$B^{0}\to {D}^{*-} K_0^{*+}(1430)$&T&$0.75_{-0.24-0.26-0.02}^{+0.26+0.28+0.05}$&$0.09_{-0.07-0.02-0.01}^{+0.14+0.02+0.01}$\\
\vspace{0.05cm}
$B^{0}\to \overline{D}^{*0} K_0^{*0}(1430)$&C&$4.90_{-1.81-0.49-0.14}^{+2.08+0.33+0.30}$&$3.80_{-2.79-0.58-0.11}^{+3.80+0.57+0.23}$\\
\vspace{0.05cm}
$B_{s}\to {D}^{*-} a_0^{+} (1450)$&E&$0.06_{-0.03-0.01-0.01}^{+0.03+0.01+0.01}$&$0.05_{-0.02-0.01-0.01}^{+0.03+0.01+0.01}$\\
\vspace{0.05cm}
$B_{s}\to \overline{D}^{*0} a_0^{0}(1450) $&E&$0.03_{-0.01-0.01-0.01}^{+0.01+0.01+0.01}$&$0.03_{-0.01-0.01-0.01}^{+0.01+0.01+0.01}$\\
\vspace{0.05cm}
$B_{s}\to \overline{D}^{*0} f_0(1370)(f_n) $&E&$0.02_{-0.01-0.01-0.01}^{+0.01+0.01+0.01}$&$0.02_{-0.01-0.01-0.01}^{+0.01+0.01+0.01}$\\
\vspace{0.05cm}
$B_{s}\to \overline{D}^{*0} f_0(1370)(f_s) $&C&$3.29_{-1.53-0.20-0.10}^{+1.94+0.14+0.20}$&$3.43_{-2.56-0.42-0.11}^{+3.50+0.32+0.20}$\\
\vspace{0.05cm}
$B_{s}\to \overline{D}^{*0} f_0(1500)(f_n) $&E&$0.02_{-0.01-0.01-0.01}^{+0.01+0.01+0.01}$&$0.02_{-0.01-0.01-0.01}^{+0.01+0.01+0.01}$\\
\vspace{0.05cm}
$B_{s}\to \overline{D}^{*0} f_0(1500)(f_s) $&C&$3.25_{-1.49-0.17-0.09}^{+1.90+0.17+0.20}$&$3.21_{-2.43-0.35-0.10}^{+3.32+0.28+0.19}$\\
\vspace{0.05cm}
$B_{s}\to D_s^{*-} K_0^{*+}(1430) $&T&$1.34_{-0.48-0.40-0.04}^{+0.52-0.42-0.08}$&$0.22_{-0.20-0.06-0.01}^{+0.28+0.05+0.01}$\\
\hline\hline
\end{tabular}\label{T4}
\end{table}

Using the obtained decay amplitudes and the input parameters above,  we tabulated the calculated branching ratios with uncertainties in tables \ref{T1}-\ref{T4}. In this work, we mainly evaluate there kinds uncertainties. The first errors are caused by hadronic parameters in the wave functions of  initial and final states mesons, such as the decay constants $f_B$, $f_S$, $\overline{f}_S$, the shape parameter $\omega_{B/B_s}$ in distribution amplitude of $B/B_s$ meson, the Gegenbauer moments $B_i$ in the distribution amplitudes of the scalar mesons; The second are from the currently unknown next-to-leading order corrections, characterized by the choice of the $\Lambda_{QCD}(0.25\pm0.05){\rm GeV}$ and the variations of the factorization scales $t$ ($0.8t\to1.2t$) in the Sudakov form factor; The last errors come form the  uncertainties of the CKM matrix elements listed in eq.(\ref{eq:parameter}). From the tables, it is apparent that the most significant theoretical uncertainties are from the hadronic parameters, because the mesons' wave functions are the most important inputs  in the PQCD approach, and heavily affect the theoretical predictions. In these tables, in order to indicate the dominant contributions, we also mark each channel  by the symbols ``T "(color-allowed tree contributions), ``C"(color-suppressed tree contributions), and ``E"($W$ exchange type contributions). All of these decays only occur through tree operators, then the $CP$ asymmetry parameters vanish in SM.

From the tables, one can find that, compared with the $\Delta S =0$ processes,  the $\Delta S =1$ processes are all suppressed by the CKM matrix elements $|V_{us}/V_{ud}|^2$. For these color-allowed (T) decays with a scalar meson emitted, the contributions from factorizable emission diagrams are either suppressed by the tiny vector decay constant of the scalar meson or even vanish for the neutral scalar mesons,  though they have large wilson coefficients. For the two hard-scattering diagrams(c and d in fig.\ref{fig:diagram2}), because the light-cone distribution amplitude $\phi_S$ of the scalar meson is  antisymmetric, the contributions no longer cancel but strengthen each other, which can be seen from eqs.(\ref{eq:enfs}) and (\ref{eq:enfsl}). So,  although the hard-scattering diagrams are suppressed by the wilson coefficient $C_1$, they also provide sizable contributions and even dominate the decay amplitudes in some decay modes.  In addition, we note that, for these color-allowed decays with a $\kappa/K_0^{*}(1430)$ emitted, the contributions from factorizable emission diagrams are still sizable, because the vector decay constant of $\kappa/K_0^{*}(1430)$ is not too small due to the SU(3) breaking.

We now discuss the color-suppressed (C) decays with a $\overline{D}^{(*)}$ meson emitted.  The factorizable emission diagrams are suppressed by the small wilson coefficient $C_1+C_2/3$. Since the cancelation between the hard-scattering emission diagrams ( c and d in fig.\ref{fig:diagram1}) is suppressed by the mass difference between the $\overline{c}$ quark and the ``light'' quark in the emitted $\overline{D}^{(*)}$ meson,  the hard-scattering emission diagrams with the large wilson coefficient $C_2$ are no longer negligible, even dominate the decay amplitudes.  Therefore, the branching ratios are expected to be large enough to be detected at ongoing experiments, especially for these $\Delta S =0$ processes.

As is known, the annihilation type diagrams are power suppressed in PQCD approach. So, the branching ratios of the pure annihilation type decays (marked by ``E'') are much smaller than the ``T'' and ``C'' type decays. But, for the $B_q \to \overline{D}^{(*)}S$ decays, because of the large mass difference between the $\overline{D}$ meson and the charmless scalar meson weakens the cancellation between the two nonfactorizable annihilation type diagrams (g) and (h) in fig.\ref{fig:diagram1}, the annihilation type contributions might be sizable. As a result, the branching ratios of these pure annihilation type decays (E) are not too small as usual, especially for these $\Delta S =0$ processes. For example, enhanced by the CKM matrix elements, the branching ratios of the $B^0 \to D_s^- K_0^{*+}(800/1430)$ even reach $10^{-5}$,  the order of which is measurable in the ongoing experiments. When the experiments are available, it will provide another platform to learn the dynamical mechanism of the annihilation diagrams in two-body hadronic $B$ decays.

The $B^{+} \to \overline{D}^{(*)0}a_0^+(980)$ and $B^+\to \overline{D}^{(*)0}\kappa^+(800)$ decays have both the $T$ type contributions with a scalar meson emitted and the $C$ type contributions with a $\overline{D}^{(*)}$ emitted. For $B^+\to \overline{D}^{(*)0} a_0^+(980)$ decays, the constructive interference between those two contributions makes the branching ratio larger than the pure $C$-type decays, such as the $B_s \to \overline{D}^{(*)0} \overline{\kappa}^0$ decay. Similarly, the constructive (destructive) interferences also preserve the branching ratio of $B^+\to \overline{D}^{(*)0}\kappa^+$ larger (smaller) than the pure ``C'' type $B^0 \to \overline{D}^{(*)0} \kappa^0$ decay. In particular, since the vector decay constant of $\kappa$ is not tiny, the $T$ type contributions with a $\kappa$ emitted are sizable.  From the Table.\ref{T1}, one can also find that $\mathcal{B}(B^0\to D^- a_0(980)^+) < \mathcal{B} (B_s \to D_s^- a_0^+ (980))$ and $\mathcal{B}(B_s \to D_s^- \kappa^+(800))<\mathcal{B}(B^0\to D^- \kappa^+(800))$, which can be understood by the destructive interferences between the emission contributions (T) and the annihilation type contributions (E).  The relation $\mathcal{B}(B^0\to D^{*-} a_0(980)^+) > \mathcal{B} (B_s \to D_s^{*-} a_0^+ (980))$ and $\mathcal{B}(B_s \to D_s^{*-} \kappa^+(800))>\mathcal{B}(B^0\to D^{*-} \kappa^+(800))$ in Table.\ref{T3} can also be attributed to the constructive interferences.

From the Tables.\ref{T2} and \ref{T4},  it is  found that, for these pure color-suppressed (C) decays, such as $B_s \to \overline{D}^{(*)0} \overline{K}_0^{*0}(1430)$ and $B^0 \to \overline{D}^{(*)0} K_0^{*0}(1430)$ decays, the branching fractions in S1 are roughly equal to those in S2. It is can be explained by the fact  that the two dominant nonfactorizable diagrams (c) and (d) in fig.\ref{fig:diagram1} will be cancelled by each other. The effects caused by the wave functions of scalar mesons are suppressed by this cancellation, and the two scenarios arrive at the roughly same branching fractions, which is similar to case of the color-suppressed $B\to D^{(*)} S$ decays in ref.\cite{ds}. We also note that, for $B^+\to\overline{D}^0 a_0^+(1450)$, the branching fraction in S2 is larger than that in S1, which is caused by the constructive interference between the color-suppressed contributions with $\overline{D}$ emitted and the color-allowed contributions with the scalar meson emitted in S2.  However, for the decay  $B^+\to\overline {D}^{*0} a_0^+(1450)$ and $B^+ \to \overline{D}^{*0} K_0^{*+}(1430)$, the destructive interferences cause their branching fractions in S2 smaller than those in S1. As for $B^+ \to \overline{D}^0 K_0^{*+}(1430)$ decay mode, although the interference between the two type contributions is also constructive in S2, the branching ratios in S2 is only sightly larger than that in S1, because the contributions from factorizable diagram  and the nonfactorizable one will be cancelled by each other, especially when the vector decay constant of $K_0^{*+}(1430)$ is no longer small as other scalar meson. We also note that the $B^+\to\overline{D}^{(*)0} a_0^+(1450)$ decays are useful to distinguish which scenario is favorable to identify the scalar mesons. Similarly, the branching fraction deferences of $B^0\to\overline{D}^{(*)0} a_0^0(1450)$ and $B^0\to\overline {D}^{(*)0} f_0(1370/1500)$ in two different scenarios are attributed to the interference between the emission contributions and the annihilation ones.

 The color-favored tree type (T) $B_s \to D_s^{(*)-} a_0^+(1450)$ decays, which are pure emission processes with a $a_0^+(1450)$ emitted, are dominated by the hard-scattering emission diagrams, since the factorizable diagrams are highly suppressed by the vector decay constant of the $a_0^+(1450)$. The ratio of branching ratios between S1 and S2 is about 2  and 4, for $B_s \to D_s^{-} a_0^+(1450)$ and $B_s \to D_s^{*-} a_0^+(1450)$, respectively. It indicates  that the branching ratios are sensitive to the  scenarios. From eq.(\ref{eq.enf}), it is found that the contributions from two hard-scattering diagrams are strengthened by each other.  When we switch S1 to S2, the changes induced by the distribution amplitudes in S2 will overlap with each other, which makes the branching ratios different from those in S1. As for $B^0 \to D^{(*)-} a_0^+(1450)$ decays, the ratio between S1 and S2 is about 4 for $B^0 \to D^{-} a_0^+(1450)$ and about 2 for $B^0 \to D^{*-} a_0^+(1450)$, which is contrary to the cases in $B_s \to D_s^{(*)-} a_0^+(1450)$ decays. This is caused by the interferences between the emission diagrams and the annihilation diagrams. In S2, the interference  is destructive for $B^0 \to D^{-} a_0^+(1450)$ decays, but constructive for $B^0 \to D^{*-} a_0^+(1450)$.  Unlike the cases of the above T-type decays with the $a_0(1450)$,  the branching ratios of the $B^0 \to D^-K_0^{*+}(1430)$ and $B_s \to D_s^-K_0^{*+}(1430)$ decays in two scenarios are roughly equal. However, for $B^0 \to D^{*-}K_0^{*+}(1430)$ and $B_s \to D_s^{*-}K_0^{*+}(1430)$ decays, the branching ratios in S1 are much larger (about 7-8 times larger) than those in S2. For the above four decays, the color-allowed factorizable emission contributions are sizable in S2, because the vector decay constant of the $K_0^{*+}(1430)$ in S2 is larger than in S1. For  $B^0 \to D^-K_0^{*+}(1430)$ and $B_s \to D_s^-K_0^{*+}(1430)$ decays, the interference between above contributions and ones of  the hard-scattering emission diagrams  is constructive,  so  their branching ratios in S2 are roughly equal to those in S1. However, this kind of interference is destructive for  $B^0 \to D^{*-}K_0^{*+}(1430)$ and $B_s \to D_s^{*-}K_0^{*+}(1430)$ decays,  and the branching ratios of them in S2 are smaller than those in S1.

Now, we turn to discuss the pure annihilation type (E) decays, which are dominated by the nonfactorizable annihilation diagrams. From Table.~\ref{T2}, one can find that the branching ratios of pure annihilation $B \to \overline {D} S$ decays in S2 are much larger than those in S1. As we know,  the cancellation between two nonfactorizable annihilation diagrams is suppressed by the large mass difference between the $b$ quark and the light quark. So,  the changes induced by the distribution amplitudes of the scalars become important,  which leads to that  the branching ratios are dependent on the scenarios obviously. Taking $B_s \to D^- a_0^+(1450)$ for illustration, the branching ratio in S2 are about three times larger than that  in S1. However, from Table.~\ref{T4}, we find that the situation is reversed for the pure annihilation $B \to \overline {D}^* S$ decays,  the discrepancies of branching ratios in different scenarios are quit small. Comparing the eq.~(\ref{af}) with eq.~(\ref{afl}), we notice that the two factorizable annihilation diagrams are cancelled by each other in $B \to \overline{D} S$ decays, but strengthened  in $B \to \overline{D}^* S$ decays. So, in $B \to \overline{D}^* S$ decays, the contributions from two factorizable diagrams are comparable with those from nonfactorizable ones. Moreover, the interference between the factorizale annihilation diagrams and the nonfactorizable ones is destructive (or constructive) in S1 (S2), which causes that the branching ratios in S2 are almost equal to or even larger than those in S1.

Although the LHCb experiment had measured the branching fractions of $B(B_s) \to \overline{D} \sigma$ and $\overline{D}f_0(980)$ \cite{lhcb}, the mixing angle $\theta$ cannot be constrained stringently due to the large uncertainties. For the sake of convenience, we presented individually the branching ratios under the pure $n\bar n$ and $s\bar s$ components in tables. Once the S1 is confirmed and the the mixing angle is fixed, one can obtained the branching ratios directly from the two predictions with $n\bar n$ and $s\bar s$ components. For instance, if the popular value ranges $[25^{\circ}, 40^{\circ}]$ and $[140^{\circ}, 165^{\circ}]$ are adopted, we can predict the branching fractions as listed in Table.~\ref{T5}. In the same manner, by neglecting the tiny glueball contents and adopting results of eq.~(\ref{eq:f02}), we also list the branching fractions of decay modes with $f_0(1370)$ or $f_0(1500)$ in Table.~\ref{T6}. Note that we here only list the center values for simplicity.

 \begin{table}[!htb]
\centering
 \caption{The calculated branching ratios of $B_{(s)}\to \overline{D}^{(*)} f_0(980)$ and $\sigma$ with the mixing in the PQCD approach (unit:$10^{-6}$).}
\begin{tabular}[t]{l!{\;\;\;\;}c!{\;\;\;\;\;\;\;}c}
\hline\hline
  \multirow{2}{*}{Decay Modes} & \multirow{2}{*}{$[25^{\circ}, 40^{\circ}]$} & \multirow{2}{*}{$[140^{\circ}, 165^{\circ}]$} \\
  &&\\
 \hline
 \vspace{0.05cm}
$B^{0}\to \overline{D}^{0}\sigma$&$78.6\sim110$&$78.6\sim125$\\
\vspace{0.05cm}
$B^{0}\to \overline{D}^{0}f_0(980)$ &$46.0\sim64.4$&$46.0\sim73.1$\\
\vspace{0.05cm}
$B_s\to \overline{D}^{0}\sigma$&$3.99\sim7.22$ &$1.66\sim5.87$\\
\vspace{0.05cm}
$B_s\to \overline{D}^{0}f_0(980)$ &$5.46\sim7.92$&$8.09\sim10.6$\\
\vspace{0.05cm}
$B^{0}\to \overline{D}^{*0}\sigma$ &$100\sim140$&$100\sim159$\\
\vspace{0.05cm}
$B^{0}\to \overline{D}^{*0}f_0(980)$&$70.2\sim98.2$&$70.2\sim111$\\
\vspace{0.05cm}
$B_{s}\to \overline{D}^{*0}\sigma$ &$2.93\sim6.70$&$1.37\sim7.09$\\
\vspace{0.05cm}
$B_{s}\to \overline{D}^{*0}f_0(980)$&$7.62\sim10.5$&$7.55\sim11.9$\\
\hline\hline
\end{tabular}\label{T5}

\end{table}
\begin{table}[!htb]
\centering
 \caption{The calculated branching ratios of $B_{(s)}\to \overline{D}^{(*)} f_0(1370)$ and $f_0(1500)$ with the mixing in the PQCD approach (unit:$10^{-6}$).}
\begin{tabular}[t]{l!{\;\;\;\;\;\;\;\;\;}c!{\;\;\;\;\;\;\;\;\;}c}
\hline\hline
  \multirow{2}{*}{Decay Modes} & \multirow{2}{*}{S1} & \multirow{2}{*}{S2} \\
  &&\\
 \hline
 \vspace{0.05cm}
$B^{0}\to \overline{D}^{0}f_0(1370)$&$173$&$99.2$\\
\vspace{0.05cm}
$B^{0}\to \overline{D}^{0}f_0(1500)$ &$79.3$&$38.5$\\
\vspace{0.05cm}
$B_s\to \overline{D}^{0}f_0(1370)$&$7.40$ &$4.85$\\
\vspace{0.05cm}
$B_s\to \overline{D}^{0}f_0(1500)$ &$14.9$&$24.5$\\
\vspace{0.05cm}
$B^{0}\to \overline{D}^{*0}f_0(1370)$ &$272$&$168$\\
\vspace{0.05cm}
$B^{0}\to \overline{D}^{*0}f_0(1500)$&$129$&$73.2$\\
\vspace{0.05cm}
$B_{s}\to \overline{D}^{*0}f_0(1370)$ &$10.7$&$7.55$\\
\vspace{0.05cm}
$B_{s}\to \overline{D}^{*0}f_0(1500)$&$20.8$&$24.6$\\
\hline\hline
\end{tabular}\label{T6}
\end{table}

In fact, under two-quark assumption, only $n\bar{n}$ component contributes to the decay modes $B^0 \to \overline{D}^0 f_0(980)$ and $B^0\to \overline{D}^0 \sigma$. Thus, we can define a ratio as
\begin{eqnarray}
r=\frac{B^0 \to \overline{D}^0 f_0(980)}{B^0\to \bar{D}^0 \sigma}=\frac{\sin^2 \theta}{\cos ^2 \theta}=\tan^2\theta.
\end{eqnarray}
Using the latest  experimental data in eq.(\ref{valueexp}), we can obtain $r=0.12^{+0.09}_{-0.06}$, which can constrain the range of mixing angle as
 \begin{eqnarray}
\theta \in [14^{\circ},24^{\circ}] ~~\textrm{or}~~ [155^{\circ},166^{\circ}].
\label{eq:newangle}
 \end{eqnarray}
 Compared with the results of ref.\cite{angle}, the obtuse angle solutions agree with each other, but the acute angle we obtained is a bit smaller than the previous results. Using the mixing angle value in eq.(\ref{eq:newangle}) and the results in Table. \ref{T1}, we get the branching ratios of $B^0 \to \overline{D} f_0(80)/\sigma$  as
\begin{eqnarray}
\mathcal{B}(B^0\to \overline{D}^0 \sigma)&\sim&11.9_{-0.8}^{+0.7} \times 10^{-5},\nonumber\\
\mathcal{B}(B^0\to \overline{D}^0 f_0(980)& \sim& 0.8_{-0.4}^{+0.5} \times 10^{-5},
\end{eqnarray}
where the errors are only from the mixing angle. Compared to eq.(\ref{valueexp}), one can find that our numerical results can accommodate the experimental data well within the limit of errors. In ref.\cite{prd93}, in order to explain the data, for the $\sigma$ meson, the authors have adopted the same decay constant and light-cone distribution amplitudes as the $a_0(980)$.

In brief, under the assumption of $q\overline{q}$ bound states of the scalar mesons,  we hope to provide a potential  way to study the substructure and physical properties of the scalar mesons, when the experiments are available, especially the LHCb and the Bell-II experiments. We also acknowledge that, for the two-body $B$ decays with a scalar meson, the nonperturbative contributions and even the exotic new physics contributions may play an important role, which have been neglected in this work because they are beyond the scope of this work, and are left for the future.

%============================================================================
\section{SUMMARY}
In this work, we investigate the $B/B_s \to \overline{D}^{(*)} S$ decays induced by $\overline{b} \to \overline{c}$ transition within the framework of the PQCD approach in two scenarios of the scalar mesons. Since the considered decays occur only through the tree operators, there are no $CP$ asymmetries. The branching ratios of the most decay modes are in the range of $10^{-4}$-$10^{-7}$, which can be tested in the LHCb experiment and the Belle-II in the near future. Some decays with large branching ratios, such as the $B^+ \to \overline{D}^{(*)0} a_0^+(980/1450)$ and $B^+ \to D^{(*)-} a_0^+(980/1450)$, which are sensitive to the scenarios, might shed light  on the structure and nature of scalar mesons. For the $B^0 \to \overline{D}^0 \sigma$ and $B^0 \to \overline{D}^0 f_0(980)$ decays, our numerical results accommodate the experimental data well within the limit of the errors.

%============================================================================
\section*{Acknowledgment}
This work was supported in part by the National Science Foundation of China under the Grant Nos.~11447032, 11575151, 11235005, 11205072, the Natural Science Foundation of Shandong province (ZR2014AQ013 and ZR2016JL001) and the Program for New Century Excellent Talents in University (NCET) by Ministry of Education of P. R. China (Grant No. NCET-13-0991), and also supported by the Research Fund of Jiangsu Normal University under Grant  No. HB2016004.
%&&&&&&&&&&&&&&&&&&&&&&&&&&&&&&&&&&&&&&&&&&&&&&&&&&&&&&&&&&&&&&&&&&&&&&&&&&7
%                     appendix
%&&&&&&&&&&&&&&&&&&&&&&&&&&&&&&&&&&&&&&&&&&&&&&&&&&&&&&&&&&&&&&&&&&&&&&&&&77

%============================================================================

\end{document}